%%%%%%%%%%%%%%%%%%%%%%%%%%%%%%%%%%%%%%%%%%%%%%%%%%%%%%%%%%%%
%%%%%%%%%%%%%%%%%%%%%%%%%%%%%%%%%%%%%%%%%%%%%%%%%%%%%%%%%%%%
%%%   A review of Johnjoe McFadden's book 
%%%       ``Quantum Evolution''.
%%%   quant-ph/0101019
%%%   
%%%
%%%   Plain TeX, 11 pages
%%%
%%%   reviewed by Matthew J. Donald
%%%
%%%   web site:  http://www.poco.phy.cam.ac.uk/~mjd1014
%%%
%%%   e-mail  :  matthew.donald@phy.cam.ac.uk
%%%%%%%%%%%%%%%%%%%%%%%%%%%%%%%%%%%%%%%%%%%%%%%%%%%%%%%%%%%%
%%%%%%%%%%%%%%%%%%%%%%%%%%%%%%%%%%%%%%%%%%%%%%%%%%%%%%%%%%%%

%%%%%%%%%%%%%%%%%%%%%%%%%%%%%%%%%%%%%%%%%%%%%%%%%%%%%%%%%%%%%%%%%
\count17=0  %to use pdfTeX comment out this line and uncomment the next
%\count17=1 \pdfoutput=\count17 %%to use plain TeX comment out this line
%%%%%%%%%%%%%%%%%%%%%%%%%%%%%%                and uncomment the previous
%%%%%%%%%%%%%%%%%%%%%%%%%%%%%%%%%%%%%%%%%%%%%%%%%%%%%%%%%%%%%%%%%

\ifnum\count17=1 

\def\cmykBlack{0 0 0 1}

\def\pdfsetcolor#1{\pdfliteral{#1 k}}

\def\maincolor{\cmykBlack}
\pdfsetcolor{\maincolor}

\def\makefootline{
    \baselineskip24pt
    \line{\pdfsetcolor{\maincolor}\the\footline}}

\def\makeheadline{%
    \edef\M{\topmark}
    \ifx\M\empty\let\M=\maincolor\fi
    \vbox to 0pt{\vskip-22.5pt
        \line{\vbox to8.5pt{}%
        \pdfsetcolor{\maincolor}\the\headline\pdfsetcolor{\M}}\vss}%
    \nointerlineskip}

\pdfpagewidth=18.0cm    
\pdfpageheight=23.4cm  

\pdfpagesattr
{ /MediaBox [0 0 612.28 795.97]
  /CropBox [50 0 570 792] }

\pdfinfo
{ /Title (A review of Johnjoe McFadden's book ``Quantum Evolution''.)
  /Author (Matthew J. Donald)
  /CreationDate (January 2001) 
  /ModDate (\number\year/\number\month/\number\day) 
  /Subject (quantum theory and biology)
		/Keywords (quantum theory, quantum evolution, McFadden)}

\pdfcompresslevel=9

%\pdfmovechars=2
\fi

\magnification=1200
\hsize=13cm
\topskip10pt plus30pt
\def\<{{<}} \def\>{{>}} 
\def\dsize{\displaystyle} 
\def\newpage{\vfill\eject}
\abovedisplayskip=3pt plus 1pt minus 1pt
\belowdisplayskip=3pt plus 1pt minus 1pt
\def\proclaim#1#2{\medskip\noindent{\bf #1}\quad \begingroup #2}
\def\endproclaim{\endgroup\medskip}
\def\proof{\noindent{\sl proof}\quad}
\def\H{{\cal H}}
\def\blacksquare{\vrule height4pt width3pt depth2pt}
\def\hcrh{\hfill \cr \hfill} \def\crh{\cr \hfill} 
\font\trm=cmr12  
\def\tila{\lower 1.1 ex\hbox{\trm \char'176}}
\def\tilb{\lower 1.6 ex\hbox{\trm\char'176}}
\def\dent{\leavevmode\hbox to 20pt{}}

\ifnum\count17=1 
\def\link#1#2{\leavevmode\pdfannotlink attr{/Border [0 0 0]} goto
name{#1}\setcolor\cmykBlue #2\pdfendlink\setcolor\cmykBlack} 
\def\name#1{\pdfdest name{#1} xyz}

\else
\def\link#1#2{{#2}}
\def\name#1{}

\fi

\centerline{BOOK REVIEW \link{*}{*}\footnote{}{\tenrm * quant-ph/0101019
\quad January 2001.}}
\medskip

\centerline{\bf Quantum Evolution} 
\centerline{\bf by Johnjoe McFadden}
\medskip

\centerline{\bf reviewed by  Matthew J. Donald} 
\medskip

{\bf \hfill The Cavendish Laboratory,  Madingley Road,  Cambridge 
CB3 0HE, 

\hfill Great Britain.}

\medskip

{\bf \hfill e-mail:\quad matthew.donald@phy.cam.ac.uk}

\medskip

{\bf \hfill web site: \quad 
{\catcode`\~=12 \catcode`\q=9 http://www.poco.phy.cam.ac.uk/q~mjd1014}}
\bigskip

\noindent J. McFadden, {\sl Quantum Evolution: Life in the Multiverse} 
(HarperCollins, 2000), 338 pp, ISBN 0-00-255948-X, 0-00-655128-9. 
A web site with a detailed summary of the book can be found at
http://www.geneticengineering.org/evolution/mcfadden.html 

\medskip

In ``Quantum Evolution'', Johnjoe McFadden makes far-reaching claims for
the importance of quantum physics in the solution of problems in
biological science.  In this review, I shall discuss the relevance of
unitary wavefunction dynamics to biological systems, analyse the
inverse quantum Zeno effect, and argue that McFadden's use of quantum
theory is deeply flawed.

In the first half of his book, McFadden both discusses the biological
problems he is interested in solving and gives an introduction to
quantum theory.  This part of the book is excellent popular science. 
It is well-written, competent, and fun.  

As far as the biology is concerned, McFadden, who is a molecular
microbiologist, has very specific, and often controversial, opinions. 
Nevertheless, he does refer to a wide range of alternative points of
view.  He certainly managed to convince me that I had swallowed too
easily the prevailing dogma (as found, for example, in chapter 1 of
\link{albert}{Albert et al.\ 1989}) about the earliest (``prebiotic'')
stage in biological evolution during which the first self-replicating
molecules appeared.  McFadden argues that this stage seems to have
happened quite fast in terms of the age of the Earth (perhaps within
100 million years) but that none of the proposed mechanisms, of which
he discusses several, give anything like a complete and plausible
picture of how to go from the early Earth's chemistry to the first
cell.  Rescuing the prevailing dogma would require a suitable sequence
of laboratory experiments, or, at the very least, plausible computer
simulations.  Unfortunately, while 100 million years is quite short in
the age of the Earth,  \name{*} it is rather long in the laboratory.

When McFadden moves on to describe quantum mechanics and its
interpretations, I feel that his touch becomes rather less sure. 
However it is hardly surprising that I can tell that he is out of his
primary field and into mine, and the outlines of his presentation still
strike me as reasonable.   Unfortunately it is the details which matter
when he attempts to apply quantum theory.  He is not sufficiently
explicit about what is going on at the level of the quantum state.  In
my opinion, this eventually leads him hopelessly astray.

As well as the appearance of self-replication, McFadden is interested
in the possibility of ``adaptive'' or ``directed'' mutation.  This is
the claim, reviewed by \link{harold}{Lenski and Mittler (1993)}, that
there is some evidence that, in some bacteria, some mutations will tend
to appear more frequently in circumstances when they are advantageous
than when they are biologically neutral.  McFadden also sketches a
version of the idea that free will is a quantum phenomenon.  In all
three cases, the technical core of McFadden's proposals involves the
quantum dynamics of molecular systems.

The analysis of quantum dynamics is not entirely straightforward. 
According to the traditional account, the wavefunctions of quantum
systems change in two quite different ways.  Sometimes there is abrupt,
indeterministic, change -- ``wavefunction collapse'' -- described by
the ``projection postulate'' in which, with a probability given by a
squared transition amplitude, the wavefunction is replaced by some
eigenfunction of a ``measurement operator''.  At all other times, the
wavefunction changes according to a Schr\"odinger equation, so that the
dynamics is defined by a unitary group of the form $U(t) = \exp( - i t
H)$ with $H$ a self-adjoint Hamiltonian operator.

More modern accounts invoke decoherence theory (\link{qtab}{Giulini et
al.\ 1996}) in order to justify the assumption that there is always some
sufficiently large scale on which the dynamics is unitary.  In
particular, decoherence theory encourages us to treat ``measurements''
as physical processes governed by appropriate global Schr\"odinger
equations and to interpret the abrupt changes as merely the way in
which changes in one part of the global quantum system would appear to
another part.

Although decoherence theory does leave open crucial conceptual problems
in the interpretation of quantum theory, it suggests that those
problems are not primarily matters of dynamics.  The Schr\"odinger
equations which apply in biological situations are precisely those of
conventional quantum chemistry and are defined ultimately by the
electromagnetic interactions between electrons and nuclei in a fixed
background classical gravitational field.   Certainly if, as he claims,
McFadden's ideas are compatible with the many-worlds interpretation of
quantum theory, then we should also be able to analyse his work in
terms of global unitary non-relativistic electromagnetic quantum
dynamics.  In general, however, subsystems of a system with unitary
dynamics will not themselves have unitary (reversible) dynamics unless
they are effectively isolated from their surroundings.  Moreover,
subsystems of a system in a pure (wavefunction) state will not
necessarily themselves occupy pure states.  Indeed, any account of
local states entirely in terms of product wavefunctions, even if it is
possible at a single instant, will disregard important issues of
thermal physics which are particularly relevant to the apparent
dynamics of biological systems.  

Consider then McFadden's proposals about the process by which
self-replicating proteins appear.  (McFadden does mention that
according to the prevailing dogma self-replication in RNA arose before
self-replication in proteins, but he argues that RNA is completely
implausible as a prebiotic chemical.)  Assume that the early Earth
produces a sea of amino acids which can link into peptides a few of
which are self-replicating.  According to McFadden, the chance of a
random peptide being self-replicating is far too small for this process
to be a plausible source of life in the framework of classical
biochemistry.  He claims that, instead, we must invoke what he calls
the ``inverse quantum Zeno effect''.  (Thorough introductions to 
quantum Zeno effects are given in chapter 8 of \link{mcfadden}{Namiki,
Pascazio, and Nakazato 1997} and in section 3.3.1 of 
\link{qtab}{Giulini et al.\ 1996}.)

The idea of the inverse quantum Zeno effect goes back to
\link{mcfadden}{von Neumann  (1932, section V.2)}. For a mathematician,
it is some version of \link{A.2}{the theorem} proved in the appendix
which shows that, up to analytical niceties, if the projection postulate
is true, and if we can choose our measurements at will, then, by using
the right dense sequences of measurements, we can turn some given pure
state
$|\Phi\>\<\Phi|$ into any other pure state $|\Psi\>\<\Psi|$  regardless
of the physical Hamiltonian.  

In order to prove this, we first need to use some form of the
mathe\-matically-simple existence lemma (\link{A.1}{A.1}) which states
that there exists a bounded self-adjoint Hamiltonian $K$ such that
$\exp(-i K) \Phi = \Psi$.  However, the problem with such an existence
lemma is that $K$ is totally artificial.  It is a purely
mathematical adjunct required to reduce to the theorem for the standard
Zeno effect in which we have  $\Psi = \Phi$.

The crucial fact on which the inverse Zeno effect depends, is that, for
any pair of unitary propagators $U(t)$ and $V(t)$ defined by suitable
Hamiltonians,
$$|\<\Phi | U(1/N) V(-1/N) | \Phi \>|^2 = 1 - O(1/N^2). \eqno(1)$$ In
\link{A.2}{theorem A.2}, this fact appears as \link{A.3}{(A.3)} with
$V(t) = e^{-itK}$.

(1) is an extension of the principle that unitary change in quantum
theory begins slowly.  Indeed, using $N \times (1/N^2) = 1/N$, the idea
is that if we can interrupt a change sufficiently often then we can
gradually alter it in any given direction.  To make use of this,
however, we actually do have to have large numbers of interruptions
applied within a bounded time.  It should also be noted that the
interruptions, or projections, constructed in the proof are strongly
dependent, through \link{A.1}{lemma A.1}, on the initial and final
wavefunctions.

Decoherence theory tells us that the projection postulate can be viewed
as a phenomenological aspect of a unitary dynamics at a large scale. 
For a preliminary version of how this works at a mathematical level, we
can use another existence lemma similar to \link{A.1}{lemma A.1}.  What
\link{A.4}{lemma A.4} means in words is that the projection postulate
can always be modelled as the restriction of a unitary dynamics on a
large space to density matrix (or mixed state) dynamics on a pair of
smaller spaces.

Once again, however, the Hamiltonian $L$ in \link{A.4}{lemma A.4} is
totally artificial.  For each of the $N$ projections $P^N_n$ in
\link{A.2}{theorem A.2} we need to have access to an auxiliary space.  We
also require that the time $s$ for which the corresponding map 
$\exp(-i s L)$ should apply should satisfy $s \sim 1/N$ and that the
dynamics should be switched off at the end of that brief period.  The
famous experiments of  \link{harold}{Itano et al.\ (1990)} involved not
only a series of brief interactions with auxiliary spaces of photon
states, but also a system in which there was an essentially
thermodynamical reason for rapid relaxation onto the projected state. 
Thus there were  very specific constraints on the form of the physical
dynamics and on the projections.

As a physical example of the inverse quantum Zeno effect, McFadden
considers three polarizing lenses.  Adjust lenses 1 and 2 to block
light entirely.  Then insert lens 3 between lenses 1 and 2.  With lens
3 turned to a suitable angle, some light will pass through the entire
sequence.

The simple analysis in the appendix is not entirely relevant to this
case, because our observations of light intensity involve the average,
macroscopic, behaviour of many photons.  Nevertheless, this is an
excellent example.  It shows that the Zeno effect is not magic.  We are
dealing here with a situation in which it possible to understand why
the projection postulate should be a reasonable model of the
interaction between the measured property of the light and the lens. 
The measurements involved are genuine and identifiable physical
processes.  Nevertheless, the required sequence of measurements has to
be carefully constructed and adjusted.  The light polarization to which
the effect is applied is a particularly simple and easily controlled
property.  Moreover, when the experiment is performed, one sees, in the
inefficiency of light transmission through a real lens, the difference
between the mathematical ideal and what is possible in practice. 

There are many ways in which evolution has become very good at careful
construction and adjustment, but this was certainly not true at the
prebiotic stage.  As a mechanism for producing desirable peptides on a
lifeless planet, my opinion is that the inverse Zeno effect has no
plausibility.  I would only change that opinion if I was presented with
a detailed model, at the level of the quantum states of all the
particles involved.  The essential difficulty is to explain why the
actual molecular dynamics should have the slightest resemblance to what
is merely a theoretically possible dynamics.

Instead of giving such an explanation, McFadden, both in his book and
in a published paper (\link{mcfadden}{McFadden and Al-Khalili 1999}) on
adaptive mutation, provides a much vaguer picture of molecular systems
as sometimes being isolated and sometimes decoherent, and he suggests
that when the systems he considers come out of isolation, suitable
measurements will drive them towards desirable biological goals.  This
picture has much in common with the idea of a quantum computer.  As
such of course we can return to the necessity for experiments.  Here
the primitiveness of our artificial quantum computers and the
difficulty that we have in building them, seems to me to be fairly good
evidence that the sort of multi-atom wavefunction control which would
be required to make McFadden's ideas work is unlikely to have occurred
by chance.

At the prebiotic stage, McFadden wants to consider a soup of as many as
20 varieties of amino acid with different side chains as being isolated
``inside tiny structures: perhaps in the pores of a rock or within a
chemically generated oil or protein droplet''.  Amino acid residues
vary in volume from 60 \AA$^3$ for glycine to 228 \AA$^3$ for
tryptophan.  They vary in structure by as many as seventeen atoms and
they vary in pH and in all sorts of other reactivities
(\link{albert}{Creighton 1983}).  I find it inconceivable that in a
peptide in a drop of warm wet fluid we could exchange a tryptophan
residue for a glycine residue without the state of every molecule in
the drop rapidly being affected.  After all, the change of volume in
the peptide would provide room for more than five water molecules.

In their analysis of adaptive mutation, McFadden and Al-Khalili
consider an entire bacterial cell as an isolated unit with a possible
mutation resulting in a protein in which an arginine residue is
replaced by a histidine residue.  In comparison to other residue pairs,
arginine and histidine are fairly similar.  They are both bases, and
they differ in volume by only about 2/3rds of a water molecule. 
Nevertheless, their structures are still substantially different, with
arginine containing one more nitrogen and five more hydrogen atoms.

McFadden and Al-Khalili produce the following equation to describe the
wavefunction of a cell in this situation:
\name{eq2}
$$|\Psi_{cell}\> = \alpha|\Phi_{not\ tun.}\>|C\>|Arg\> +
\beta|\Phi_{tun.}\>|T\>|His\>.
\eqno{(2)}$$ Here $\Phi$ is supposed to denote the wavefunction of a
proton on a gene which can tunnel between two possible positions. 
$|C\>$ and
$|T\>$ denote corresponding wavefunctions of a daughter DNA strand
following replication and $|Arg\>$ and $|His\>$ denote the wild-type
and mutant form of the protein.

At best, equation \link{eq2}{(2)} can be considered as schematic,
ignoring as it does all the other particles which will be affected by the
processes of DNA replication and transcription and protein synthesis. 
Indeed, as they refer to different numbers of particles, neither
``$|C\>$'' and ``$|T\>$'' nor ``$|Arg\>$'' and ``$|His\>$'' can even be
considered to belong to the same Hilbert spaces.  This is not just poor
notation; except in special circumstances like superconductivity, the
more particles are involved in a situation, the less plausible coherence
becomes.  As I shall discuss further below, there is no justification
for the assignment of a pure state over an extended period to a system
like a bacterial cell.  This difficulty will extend even to quite small
parts of the cell, including single proteins and nucleic acids with their
complicated internal thermal mobilities (\link{harold}{McMammon and
Harvey 1987}).  It will certainly extend to the groups of molecules
necessarily involved in the two components of equation \link{eq2}{(2)}. 
As a general rule, coherence is destroyed at least as easily as the most
minute quantities of heat are passed.  In a fluid system, heat is
rapidly exchanged not only between the translational degrees of freedom
of separate molecules but also between internal rotational and
vibrational degrees of freedom.

McFadden and Al-Khalili suppose that, in the absence of a substrate on
which the cell can grow, the coherence expressed by equation
\link{eq2}{(2)} can persist over times (1 -- 100 s) of biological
significance.  They attempt to justify this claim by refering to
measurements of NMR $T_1$ spin-lattice relaxation times for protons in
biological systems.  This is an error not only because equation
\link{eq2}{(2)} involves far more extensive state changes than the motion
of a single proton, but also because it is much easier for heat to pass
between position degrees of freedom than to or from nuclear spins.  They
then suppose that if the cell is provided with a substrate for which the
mutated protein ``His'' is adaptive, the probability of the second term
in the wavefunction will somehow be enhanced.  They appear to claim that
this will happen simply because, in the presence of an adaptive
substrate, the decoherence time will decrease.  Unfortunately, because
``His'' is the mutated and rarer form of the protein, we should assume
in normal circumstances that
$|\beta| < |\alpha|$ and that, for short times $t$, at least in the
simplest model, $|\beta|^2 \sim \sin^2 \lambda t \sim \lambda^2 t^2$. 
This makes mutation analogous to decay of a quantum energy level.  But
the normal idea is that ``observation'', or at least the standard
quantum Zeno effect, should slow down decay (albeit, usually to an
imperceptibly small extent) rather than speed it up.  Thus McFadden and
Al-Khalili really do require something like the inverse quantum Zeno
effect with its precisely adjusted interactions. Although such
interactions might exist in principle at a mathematical level, I see no
reason to believe that biological components can provide them; either
spontaneously, or by natural evolution, or by human construction.

Quantum theory is undoubtedly important for the complete understanding
of the dynamics of biological systems for reasons which go somewhat
beyond the relevance of quantum chemistry in all detailed studies of
molecular interactions.  In particular, some electron transfer systems,
such as those involved in photosynthesis, can only be properly
described by a detailed quantum-statistical-mechanical analysis
(\link{albert}{DeVault 1984}, \link{harold}{Kilin 2000}). 
Photosynthesis itself demonstrates that it is possible for biological
systems to evolve to capture individual photons.  The resulting systems
are wonderfully complex and it seems that they could only have arisen
after a long history of refinement of membrane-bound electron transfer
systems (\link{albert}{Albert et al.~1989, chapter 7}).  Even so,
excitation of an electron on one molecule, and transfer of that
electron to a succession of different molecules involves the ordinary
building blocks of biology; collisions between specific molecules in
the right environmental conditions.  The execution of an arbitrary
sequence of mimics of the projection postulate would require an
entirely different level of wavefunction control.

\link{mcfadden}{Ogryzko (1997)} also attempts to explain adaptive
mutation in quantum mechanical terms.  His suggestion is that some kind
of environment-dependent measurement process is involved.  Like
McFadden and Al-Khalili, Ogryzko attempts to extend a physical theory,
appropriate for some simple systems, way beyond its normal domain of
application.  Here too, my opinion is that the proposal has no
plausibility, and challenge Ogryzko to produce an analysis which is
explicit about the form of his ``measurement operators''.

McFadden mentions the anthropic principle as an alternative way of
dealing with the apparent improbability of the first step in the
evolution of life.  The argument here is that an event of any non-zero
probability, however low, will occur somewhere within the
``multiverse'' of many-worlds theory.  Thus we might observe a past in
which a self-replicator came into being in a random sequence of peptide
linkages, and we inevitably will observe such a past, if it is the only
way to explain our existence as observers.  McFadden rejects this
argument, both because he ``would dearly like to believe that we are
not alone in the universe'', and because, as he points out, the
argument fails to explain the crucial issue, which is how
self-replicators can have plausibly arisen within such an early period
of the Earth's existence.  If the only way that observers could exist
was through an extremely unlikely quantum event followed by biological
evolution, then we would expect our universe to have lots of planets;
so that it would be very big (which it may well be), and/or very old
(which it probably is not).  We would also expect that, when the
quantum event happened, it would happen at a typical time within the
life of the planet on which it happened, as long as there was still
time for the required subsequent evolution.  As the expected main
sequence life of the Sun is around ten billion years, while evolution
has occured within four billion years, we would therefore expect to see
geological evidence of a several billion year period during which the
Earth was lifeless.  On the other hand, if we were prepared to make the
argument using self-replicators from space, then we would expect our
observed universe to have lived for far more star lifetimes than it
apparently has.   

In the final chapter of his book, McFadden turns his attention to the
brain.  Here he suggests that the neural electromagnetic field might be
the seat of consciousness.  Personally, I fail to understand why he, or
anyone else, should believe that such a spatially-extended field can
solve the problem of the unity of consciousness in any way that a
spatially-extended pattern of neural firings cannot.  He  confuses the
quasi-classical electro-magnetic field with the wave function of the
quantization of that field.  He also speaks of single photon events in
the circumstances of voltage-gated ion channel opening in which there
is no reason to suppose that the state moves between eigenstates of a
photon number operator.

In my opinion, the most fundamental mistake in McFadden's work is the
assumption that unitary wavefunction dynamics can provide an accurate
description of the behaviour of systems such as entire cells over
significant time periods.  The dynamics implied by this assumption is
incorrect under any interpretation of quantum mechanics.  An
interpretation in which a cell did have a wavefunction at all times
would require extremely frequent wavefunction collapse, contradicting
unitarity.  

Detailed arguments against mistakes of this kind usually involve models
of the effect of scattering processes on quantum states
(\link{harold}{Joos and Zeh 1985}, \link{qtab}{Giulini et al.\ 1996
Section 3.2},
\link{mcfadden}{Tegmark 2000}).  Defence against such arguments require
analysis of lengthscales and timescales, of the details of interactions,
and of the type of quantum system within which coherence is claimed
(\link{qtab}{Hagan,  Hameroff, and Tuszynski 2000}).  At the micron
scale of a bacterial cell, coherence is only observed in very special
systems such as lasing light and superconducting electrons.  In the
first case, we have self-reinforcement (or ``amplification'') of the
macroscopic quantum properties, and in the second, we have a
multiply-occupied ground state isolated by an energy gap.  In both
cases, as in the case of light polarization considered earlier, the
macroscopic nature of the situation is neutralized by macroscopic
uniformity and by the simplicity of the interactions of certain degrees
of freedom.

Careful isolation and state preparation are the hallmarks of
experimental science and can be found in most cases for which the
projection postulate provides a good model of the ultimate
observations.  A fascinating recent example comes from the diffraction
of fuller\-ene molecules (\link{albert}{Arndt et al.\ 1999}).  In this
case, coherence is observed in the centre-of-mass of hot complex
molecules.  State preparation is by careful collimation.  Isolation,
both from internal and external influence, is for a short flight in an
evacuated chamber.  Controlled state preparation for a system as
complex as a bacterial cell is essentially impossible, while in order
to produce the required isolation over the required period, the cell
would have to be placed in a vacuum which would be fatal to it.

On the other hand, there are also examples in which entirely natural
systems do exhibit unitary wavefunction dynamics over indefinite
periods.  For example, there can be no doubt that it is appropriate to
assign a stationary pure state to the core electrons of an atom inside
a biological system.  Here the isolation is due to the large quantity
of energy which would have to be supplied in order to free the
electrons.  The state preparation is thermodynamical.  The nucleus
would have been unable to trap the electrons had they not lost energy
to the electromagnetic field.  

Beyond these various special systems, it is possible that wavefunction
dynamics may be found at the level of the entire universe.  Otherwise,
we are always entitled to require detailed and specific models and
explanations of why entropy and temperature are not relevant parameters
in the observed behaviour.  Biological systems, in particular, are
fundamentally warm and wet.  Of course cells are precisely structured
at the level of the compartments into which they divide their various
functions.  However, at the molecular level within those compartments,
or on the membranes which bound them, heat not only keeps life moving,
but also provides energy to overcome activation barriers and
dissipation to maintain irreversibility.  Rotating and vibrating,
molecules roll and rock from collision to collision until, when they
find the right partners and the right enzyme, they find a free energy
gradient down which they can fall (\link{harold}{Harold 1986}).   The
partners are transformed and their successors wander off for new
adventures. 

A correct quantum theoretical analysis of biological systems has to take
account of the local free energy gradients which drive them.  This
implies that our descriptions of most of the degrees of freedom
relevant to the functioning of biological systems should involve mixed
states and chance.  In many ways this simplifies the description of the
dynamical processes involved; classical biochemistry provides a precise
picture of the typical dynamics.  Nevertheless, at some point, we need
to investigate the fundamental meaning of quantum theoretical chances
and descriptions.  This brings us back to the crucial open conceptual
questions of decoherence theory.  My own opinion (\link{qtab}{Donald
1990, 1999}) is that the study of biological systems may be no less
important in answering these questions than is the study of quantum
theory for a complete understanding of the observed dynamics of
biological systems. 
\medskip

\proclaim{Acknowledgement}{}  The genesis of this review was a question
on the newsgroup sci.physics.research from Des Grenfell.  It is a
pleasure to thank him and other contributors to the ensuing thread;
including Charles Francis, Greg Egan, Tim,  ``fuzzy thinking'',
``toady'' (Jim), and ``trex'', for introducing me to the book, for
their suggestions, and for their comments on a preliminary version of
this review.  I should also like to thank the moderators of the
newsgroup for the work they do in keeping sci.physics.research an
interesting venue for discussion.
\endproclaim

\newpage

\proclaim{Appendix}{}
\endproclaim

\name{A.1}
\proclaim{lemma A.1}{\sl} Let $\H$ be a Hilbert space $\H$ and
$\Phi, \Psi \in \H$ with $||\Phi|| = ||\Psi|| =1$.

Then there exists a bounded self-adjoint operator $K$ on $\H$ such that 
$e^{-i K} \Phi =\Psi$.
\endproclaim

\proof  Define $\delta \in [0, 2\pi)$  by $\<\Phi| \Psi\> =
e^{-i\delta}|\<\Phi| \Psi\>|$ with
$\delta = 0$ if $\<\Phi| \Psi\> = 0$.

If $|\<\Phi| \Psi\>| = 1$ then $\Psi = e^{-i\delta} \Phi$, so choose $K
=
\delta$.  

Otherwise $|\<\Phi| \Psi\>| < 1$ and
$\dsize \Phi_\perp = { e^{i\delta}\Psi - |\<\Phi| \Psi\>| \Phi \over
\sqrt{1 - |\<\Phi| \Psi\>|^2}}$ satisfies $\<\Phi| \Phi_\perp\> = 0$
and 
$\<\Phi_\perp| \Phi_\perp\> = 1$.

Set $\sin \theta = \sqrt{1 - |\<\Phi| \Psi\>|^2}$ with $0 \leq \theta
\leq {\pi\over 2}$. Then $\cos \theta = |\<\Phi| \Psi\>|$ and so 
$e^{i\delta}\Psi = \Phi \cos \theta + \Phi_\perp \sin \theta$.

Now let $K' = - i\theta |\Phi\>\<\Phi_\perp| + i\theta
|\Phi_\perp\>\<\Phi|$ and $K = K' + \delta$.

$-iK' = -\theta |\Phi\>\<\Phi_\perp| + \theta |\Phi_\perp\>\<\Phi|$ and
$(-iK')^2 = -\theta^2 (|\Phi\>\<\Phi| + |\Phi_\perp\>\<\Phi_\perp|)$.

Thus $(-iK')^{2N} = (-1)^N \theta^{2N} (|\Phi\>\<\Phi| +
|\Phi_\perp\>\<\Phi_\perp|)$ 

and
$(-iK')^{2N+1} = (-1)^N \theta^{2N+1}(-|\Phi\>\<\Phi_\perp| +
|\Phi_\perp\>\<\Phi|)$ and so
$$\displaylines{ \hfill
 e^{-i K'} \Phi = \sum_{N\geq0} {(-iK')^{N} \over N!} \Phi = 
\Phi \cos \theta + \Phi_\perp \sin \theta = e^{i\delta}\Psi. \hfill
\blacksquare   }$$

\name{A.2}
\proclaim{theorem A.2}{\sl} Let $H$ be a self-adjoint operator on a
Hilbert space $\H$.  Let $Q_m$ be the spectral projection of $H$ for
the interval
$[-m, m]$.  Suppose $\Phi, \Psi \in \cup_{m = 1}^\infty Q_m\H$ are
normalized.  Write
$U(t) = e^{-itH}$.

Then there exist sequences $(P^N_n)_{n = 1}^{N}$ of projections such
that,
$$\displaylines{
 |\Psi_N\>\<\Psi_N| \rightarrow |\Psi\>\<\Psi| \hbox{ as } N \rightarrow
\infty \cr \hbox{ where }
\Psi_N = P_N^N U(1/N) P^N_{N-1} U(1/N) \dots U(1/N) P^N_2 U(1/N) P^N_1
U(1/N) \Phi.    }$$
\endproclaim

\proof  Let $K$ be the bounded self-adjoint operator constructed in 
lemma A.1 with $e^{-i K} \Phi = \Psi$.  Note that we can choose $M$
such that
$\Phi, \Psi \in Q_M \H$, and then, from the proof of lemma A.1, for all
$t$,
$e^{-i t K}\Phi \in Q_M \H$.

Define $P_n^N =  e^{-i n K /N } |\Phi\>\<\Phi| e^{i n K /N }$.

As $||\Psi|| = 1$ and $||\Psi_N|| \leq 1$, to prove that
$|\Psi_N\>\<\Psi_N|
\rightarrow |\Psi\>\<\Psi|$, it is sufficient to prove that $|\<\Psi |
\Psi_N\>|^2 \rightarrow 1$ or that
$\log |\<\Psi |
\Psi_N\>|^2 \rightarrow 0$.  
$$|\<\Psi | \Psi_N\>|^2 = \prod_{n =1}^N |\<\Phi| e^{i n K /N } e^{- i
H/N} e^{-i (n-1) K/N} |\Phi\>|^2.$$

Let $|\Phi_n\> = e^{-i n K/N} |\Phi\>$ and
$$ a_n(t) = |\<\Phi_n| e^{- i t H}  e^{i t K}|\Phi_n\>|^2
 = \<\Phi_n| e^{-i t K} e^{ i t H} |\Phi_n\> \<\Phi_n| e^{- i t H} 
e^{i t K} |\Phi_n\>. $$

$a_n(t)$ is infinitely continuously differentiable.  $a_n(0) = 1$, 
$$\displaylines{
 a_n'(t) = i(\<\Phi_n| e^{-i t K} (H - K) e^{ i t H} |\Phi_n\> \<\Phi_n|
e^{- i t H}  e^{i t K} |\Phi_n\>
\hcrh + \<\Phi_n| e^{-i t K} e^{ i t H} |\Phi_n\> \<\Phi_n| e^{- i t H}
(K - H)  e^{i t K} |\Phi_n\>), }$$
$a_n'(0) = 0$, 
$$\displaylines{
 a_n''(t) = -(\<\Phi_n| e^{-i t K} (H^2 - 2KH + K^2) e^{ i t H}
|\Phi_n\>
\<\Phi_n| e^{- i t H}  e^{i t K} |\Phi_n\>
\hcrh + 2 \<\Phi_n| e^{-i t K} (H - K) e^{ i t H} |\Phi_n\> \<\Phi_n|
e^{- i t H} (K - H)  e^{i t K} |\Phi_n\>
\crh + \<\Phi_n| e^{-i t K} e^{ i t H} |\Phi_n\> \<\Phi_n| e^{- i t H}
(K^2 - 2HK + H^2)  e^{i t K} |\Phi_n\>), }$$ and
$|a_n''(t)| \leq  4(M + ||K||)^2$.

\name{A.3}
Thus 
$$a_n(t) = 1 + \int_0^t a_n'(u) du = 1 + \int_0^t \int_0^u a_n''(v) dv
du$$ and so
$|a_n(t) - 1| \leq 2 t^2 (M + ||K||)^2$. \hfill (A.3)

For $ {1\over 2} \leq x \leq 1$, $0 \leq -\log x = \dsize\int_x^1 {1
\over y} dy \leq 2(1 - x)$ and so, for $N$ sufficiently large,
$$\displaylines{
 0 \leq -2\log |\<\Psi_N | \Psi\>| = -\sum_{n =1}^N \log a_n(1/N)  \leq
\sum_{n =1}^N 4 (1/N)^2 (M + ||K||)^2 \hcrh = 4(M + ||K||)^2/N. \hfill
\blacksquare   }$$

\name{A.4}
\proclaim{lemma A.4}{\sl} Let $P$ be a projection on a Hilbert space
$\H$.  Let $\Phi \in \H$ with $||\Phi|| = 1$.  Choose $s > 0$.  Suppose
that $\H'$ is an auxiliary Hilbert space with $\Psi_0, \Psi_1, \Psi_2
\in \H'$ normalized and
$\Psi_1$ and $\Psi_2$ orthogonal.  

Then there exists a bounded self-adjoint Hamiltonian $L$ on $\H \otimes
\H'$  such that
$$\exp(-i s L)(\Phi \otimes \Psi_0) = P \Phi \otimes \Psi_1  + 
(1-P)\Phi
\otimes
\Psi_2.$$

The resulting change of state on $\H$ is $$|\Phi\>\<\Phi| \rightarrow
P|\Phi\>\<\Phi|P + (1-P)|\Phi\>\<\Phi|(1-P)$$ and on $\H'$ is
$$|\Psi_0\>\<\Psi_0| \rightarrow ||P\Phi||^2 |\Psi_1\>\<\Psi_1| +
||(1-P)\Phi||^2 |\Psi_2\>\<\Psi_2|.$$
\endproclaim

\proof  As
$||P \Phi \otimes \Psi_1  +  (1-P)\Phi \otimes \Psi_2|| = 1$, this
lemma is an immediate consequence of \link{A.1}{lemma A.1}. \hfill
$\blacksquare$
\medskip

\newpage

\proclaim{References}{}
\endproclaim

\frenchspacing
\parindent=0pt

{\everypar={\hangindent=0.75cm \hangafter=1} 

\name{albert}  Albert, B., Bray, D., Lewis, J., Raff, M., Roberts, K.,
and Watson, J.D. (1989) {\sl Molecular Biology of the Cell\ } 2nd
Edition  (Garland).

Arndt, M., Nairz, O., Voss-Andreae, J., Keller, C.,  van der Zouw, G.,
and  Zeilinger, A. (October 1999) ``Wave-particle duality of C$_{60}$
molecules.''  {\sl Nature \bf 401}, 680--682.  There is also a
first-rate web page about this experiment at  

\qquad \qquad http://www.quantum.univie.ac.at/research/c60/

Creighton, T.E. (1983) {\sl  Proteins} (Freeman).  

DeVault, D. (1984) {\sl  Quantum-Mechanical Tunnelling in Biological
Systems
\ } 2nd Edition (Cambridge).  

\name{qtab} Donald, M.J. (1990) ``Quantum theory and the brain.''  {\sl
Proc. R. Soc. Lond. \bf A 427},  43--93 and 
{\catcode`\~=12 \catcode`\q=9 http://www.poco.phy.cam.ac.uk/q~mjd1014}

Donald, M.J. (1999)  ``Progress in a many-minds interpretation of
quantum theory.'' {\sl  quant-ph/9904001} and
{\catcode`\~=12 \catcode`\q=9 http://www.poco.phy.cam.ac.uk/q~mjd1014}

Giulini, D., Joos, E., Kiefer, C., Kupsch, J., Stamatescu, I.-O., and
Zeh, H.D. (1996) {\sl Decoherence and the Appearance of a Classical
World in Quantum Theory}  (Springer).

Hagan, S., Hameroff, S.R., and Tuszy\'nski, J.A. (2000) ``Quantum
computation in brain microtubules?  Decoherence and biological
feasibility.'' {\sl quant-ph/0005025}

\name{harold} Harold, F.M. (1986) {\sl  The Vital Force: A Study of
Bioenergetics} (Freeman).  

Itano, W.M., Heinzen, D.J., Bollinger, J.J., and Wineland, D.J. (1990) 
``Quantum Zeno effect.''  {\sl Phys. Rev. \bf A 41}, 2295--2300.

Joos, E. and Zeh, H.D. (1985) ``The emergence of classical properties
through interaction with the environment.'' {\sl Z. Phys. \bf B 59},
223--243.

Kilin, D.S. (2000) ``The role of the environment in molecular
systems.''  Ph.D. Thesis, Technische Universit\"at, Chemnitz.  {\sl
quant-ph/0001004}

Lenski, R. E. and Mittler, J.E. (1993) ``The directed mutation
controversy and neo-Darwinism.'' {\sl Science  \bf 259}, 188--194.

McCammon, J.A. and Harvey, S.C. (1987) {\sl  Dynamics of Proteins and
Nucleic Acids} (Cambridge).

\name{mcfadden} McFadden, J. and Al-Khalili, J. (1999) ``A quantum
mechanical model of adaptive mutation.''  {\sl Biosystems \bf 50},
203--211.

Namiki, M., Pascazio, S., and Nakazato, H.  (1997)  {\sl  Decoherence
and Quantum Measurements} (World Scientific).

von Neumann, J. (1932)  {\sl Mathematische Grundlagen der
Quantenmechanik} (Springer) English translation (1955) {\sl
Mathematical Foundations of Quantum Mechanics} (Princeton).

Ogryzko, V.V. (1997)  ``A quantum-theoretical approach to the
phenomenon of directed mutations in bacteria (hypothesis).'' {\sl
Biosystems \bf 43}, 83--95.

Tegmark, M. (2000)  ``The importance of quantum decoherence in brain
processes.'' {\sl Phys. Rev. \bf E 61}, 4194--4206.  {\sl
quant-ph/9907009}

}

\end